\documentclass[aps,twocolumn,letter,showpacs,preprintnumbers,amsmath,amssymb,superscriptaddress,bibnotes]{revtex4}

\usepackage{graphicx,epsfig}
\usepackage{dcolumn}
\usepackage{bm} 

\begin{document}

\title{Angle-resolved photoemission studies of lattice polaron formation in the cuprate Ca$_{2}$CuO$_{2}$Cl$_{2}$}

\author{K. M. Shen}
 \email{kmshen@ccmr.cornell.edu}
 \affiliation{Department of Applied Physics, Department of Physics, and Stanford Synchrotron Radiation Laboratory, Stanford University, Stanford,
 California 94305}
 \affiliation{Department of Physics and Astronomy, University of British Columbia, Vancouver, British Columbia, V6T 1Z4, Canada}

\author{F. Ronning}
 \altaffiliation[Present address: ]{Los Alamos National Laboratory, Los Alamos, NM 87545, USA}
 \affiliation{Department of Applied Physics, Department of Physics, and Stanford Synchrotron Radiation Laboratory, Stanford University, Stanford,
 California 94305}

\author{W. Meevasana}
 \affiliation{Department of Applied Physics, Department of Physics, and Stanford Synchrotron Radiation Laboratory, Stanford University, Stanford,
 California 94305}

\author{D.H. Lu}
 \affiliation{Department of Applied Physics, Department of Physics, and Stanford Synchrotron Radiation Laboratory, Stanford University, Stanford,
 California 94305}

\author{N.J.C. Ingle}
 \affiliation{Department of Applied Physics, Department of Physics, and Stanford Synchrotron Radiation Laboratory, Stanford University, Stanford,
 California 94305}
 \affiliation{Department of Physics and Astronomy, University of British Columbia, Vancouver, British Columbia, V6T 1Z4, Canada}

\author{F. Baumberger}
 \altaffiliation[Present address: ]{School of Physics and Astronomy, University of St. Andrews, St. Andrews, Scotland}
 \affiliation{Department of Applied Physics, Department of Physics, and Stanford Synchrotron Radiation Laboratory, Stanford University, Stanford,
 California 94305}

\author{W.S. Lee}
 \affiliation{Department of Applied Physics, Department of Physics, and Stanford Synchrotron Radiation Laboratory, Stanford University, Stanford,
 California 94305}

\author{L.L. Miller}
 \affiliation{Department of Chemistry, University of Oregon, Eugene, Oregon 97403}

\author{Y. Kohsaka}
 \affiliation{Department of Advanced Materials Science, University of Tokyo, Kashiwa, Chiba 277-8561, Japan}

\author{M. Azuma}
 \affiliation{Institute for Chemical Research, Kyoto University, Uji, Kyoto 611-0011, Japan}

\author{M. Takano}
 \affiliation{Institute for Chemical Research, Kyoto University, Uji, Kyoto 611-0011, Japan}

\author{H. Takagi}
 \affiliation{Department of Advanced Materials Science, University of Tokyo, Kashiwa, Chiba 277-8561, Japan}
 \affiliation{RIKEN, The Institute for Physical and Chemical Research, Wako 351-0198, Japan}

\author{Z.-X. Shen}
 \affiliation{Department of Applied Physics, Department of Physics, and Stanford Synchrotron Radiation Laboratory, Stanford University, Stanford,
 California 94305}

\begin{abstract}

To elucidate the nature of the single-particle excitations in the undoped parent cuprates, we have performed a detailed study of Ca$_{2}$CuO$_{2}$Cl$_{2}$ using photoemission spectroscopy. The photoemission lineshapes of the lower Hubbard band are found to be well-described by a polaron model. By comparing the lineshape and temperature dependence of the lower Hubbard band with additional O $2p$ and Ca $3p$ states, we conclude that the dominant broadening mechanism arises from the interaction between the photohole and the lattice. The strength of this interaction was observed to be strongly anisotropic and may have important implications for the momentum dependence of the first doped hole states.\end{abstract}

\pacs{74.20.Rp, 74.25.Jb, 74.72.-h, 79.60.-i\\ {\bf Published in Physical Review B 75, 075115}}


\maketitle

One of the most powerful aspects of angle-resolved photoemission spectroscopy (ARPES) is the potential to quantitatively compare experimental lineshapes with theoretical predictions, providing great insight into the nature of the single particle excitations \cite{Damascelli03}. For instance, good agreement between ARPES and theory has been demonstrated on model systems such as noble metal surface states \cite{Reinert01}, two-dimensional Fermi liquids \cite{Claessen92}, and systems with strong electron-phonon coupling \cite{Hengsberger99}. Extending this quantitative analysis to the high-T$_{\mathrm{c}}$ cuprate superconductors has yielded a wealth of information regarding the salient many-body interactions \cite{Damascelli03,Valla99}. In principle, ARPES studies of the undoped Mott insulator should provide the ideal starting point to study the interactions in the cuprates, yet paradoxically, a logically consistent understanding of the spectra from the insulator has proven far more elusive than the doped compounds. However, an understanding of this simplest case is clearly essential before one can address the doped cuprates in a systematic manner. 

Although calculations based around the $t-J$ model have successfully approximated the peak dispersion in the parent insulators \cite{Dagotto94,Tohyama00}, they have been unable to explain the broad lineshapes. Attempts to model spectra from the insulator within the same weak-coupling framework used for the optimally doped cuprates have proven unsuccessful. Very recently, a phenomenological picture based on Franck-Condon broadening (FCB) was proposed to explain these features. In this scenario, the broad peak was not a coherent quasiparticle (QP), but a manifold of states with the photohole strongly coupled to bosonic shake-off excitations. Within this model, previously unresolved issues such as the unusually broad width, the lineshape, and the large separation of the peak from the chemical potential, $\mu$, have been explained in a self-consistent manner \cite{Shen04}. However, the microscopic origin of the broadening itself remains a fundamental but unresolved issue. In the past, most theoretical works considered only the effects of electronic correlations (i.e. Hubbard or $t-J$ models). Recent work has indicated that electron-phonon interactions may also be significant, and thus it is important to determine the contribution of lattice interactions. Through a comparison with conventional ``benchmark'' states (O $2p_{\pi}$ and Ca $3p$ orbitals), with the complex and strongly correlated states of the lower Hubbard band (LHB), we conclude that lattice effects play a substantial role in the observed broadening, implying that the photoholes form localized small polaron states. In addition, we have observed a  large momentum dependence to the linewidth which may have implications on the anisotropy of the first doped QP states in the lightly doped cuprates.

ARPES measurements were performed with a Scienta SES-200 electron analyzer at Beamline \mbox{5-4} of the Stanford Synchrotron Radiation Laboratory, as well as with a Scienta SES-2002 analyzer and a He plasma discharge lamp. Data were collected with photon energies between 14-32 eV at the synchrotron, and both the He I (21.22 eV) and He II (40.8 eV) lines of the discharge lamp. Typical energy and angular resolutions were better than 15 meV and 0.35$^{\circ}$, respectively. Multiple batches of Ca$_{2}$CuO$_{2}$Cl$_{2}$ were prepared using both a high pressure flux method \cite{Kohsaka02} and an ambient pressure technique \cite{Miller90}.

\begin{figure}[t!]
\includegraphics{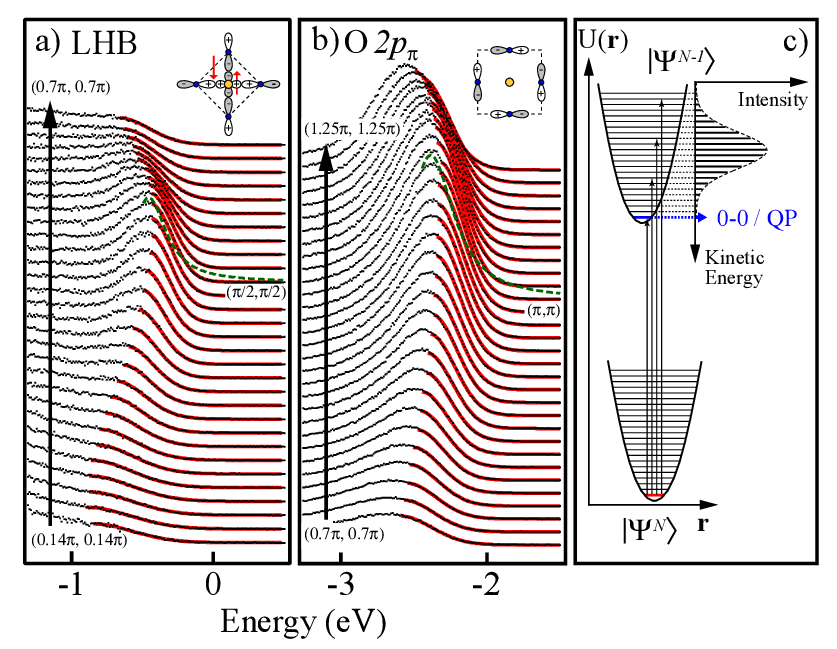} \vspace{0in}
\caption{(a) The LHB and (b) O $2p_{\pi}$ bands along $(0,0)-(\pi,\pi)$ taken with $h\nu$ = 25.5 eV and T = 300 K, together with Gaussian fits (red) and orbital configurations (top). Lorentzian fits to the peaks at  $(\pi/2,\pi/2)$ and $(\pi,\pi)$ are shown in dashed green. (c) Schematic of the Franck-Condon process, with the transitions from the initial state $| \Psi^{N}_{0} \rangle$ to possible final states $| \Psi^{N-1}_{m} \rangle$ resulting in the observed photoemission intensity.} \label{EDCCuts}
\end{figure}

In Fig. \ref{EDCCuts}a, we show energy distribution curves (EDCs) from Ca$_{2}$CuO$_{2}$Cl$_{2}$ taken near $\mathbf{k}$ = $(\pi / 2,\pi / 2)$ showing the LHB excitations. The chemical potential, $\mu$, was determined from a Au reference in electrical contact with the sample. $\mu$ was found to be pinned within the charge-transfer gap by surface defects or impurities and varied from cleave to cleave. From a large series of measurements, we found that $\mu$ was always separated from the LHB peak by at least 0.42 eV, and here we reference $\mu$ using this minimum value. Even at the top of the valence band, $\Gamma_{\mathrm{LHB}}$ is extremely broad, while simple electron scattering phase space constraints would predict extremely sharp and long lived QP excitations. In addition to the width, we also show in Fig. \ref{EDCCuts}a that the raw data can be fitted extremely well with a pure Gaussian function (with no additional parameters or background subtraction) over a wide range in both momentum and energy. At $(\pi / 2,\pi / 2)$, we also include a fit of the data to a pure Lorentzian lineshape (green dashed line) to illustrate the clear discrepancy between the experimental data and the Lorentzian fit (which should closely approximate the expectations from Fermi liquid theory near Fermi energy or at the top of the valence band). While the Gaussian lineshape is a much better fit to the data, it is inconsistent with a simple coherent single-particle spectral function, $\mathcal{A}(\mathbf{k},\omega)$ \cite{Damascelli03}. On the other hand, a Gaussian lineshape is naturally expected for the simplest case of FCB in the limit of strong coupling, where the intensity is known to follow a Poisson distribution \cite{Mahan00}.

In Fig. \ref{EDCCuts}c, we show a schematic of the FCB process, where the initial state has finite transition probabilities to many closely spaced final states, and only the lowest transition would correspond to the QP. In principle, coupling of the electrons to any bosonic field could cause FCB, so studying the LHB alone is insufficient to discriminate between phonons or magnons. Therefore, we have compared the LHB with other states which are noninteracting with the spin system, but which necessarily inhabit the same lattice. We compare the LHB with the O $2p_{\pi}$ band, shown in Fig. \ref{EDCCuts}b, as first proposed by Pothuizen \emph{et al.} in Ref. \onlinecite{Pothuizen97}. At $\mathbf{k}$ = $(\pi,\pi)$ the O $2p_{\pi}$ band is non-bonding with any Cu $3d$ states or the Zhang-Rice singlet state \cite{Pothuizen97, Hayn99} and thus is decoupled from the spin dynamics. For both the LHB and O $2p_{\pi}$, we fit only the low binding energy half of the peak. This is due to the fact that both the LHB and O $2p_{\pi}$ states have neighboring states at higher binding energies, as shown by the cluster calculations of Eskes \emph{et al.} \cite{Eskes90} and band structure calculations \cite{Hayn99}. Therefore, the spectral weight at higher energies ($<$ -0.7 eV for the LHB and $<$ -2.5 eV for the O $2p_{\pi}$ band) contains contributions from multiple bands and therefore cannot be easily fit to a single function. In our analysis, we are primarily concerned with the rapid suppression of the tail of spectral weight when going towards lower energies (closer to E$_{\mathrm{F}}$), a hallmark of FCB, and which should not be strongly affected by higher energy states or background. 

The good agreement of the Gaussian fits with not only the LHB, but also the O $2p_{\pi}$ states indicates that a FCB lineshape is generic to \emph{all} electronic states in the crystal. This would suggest that the photoholes form localized lattice polarons, since the O $2p_{\pi}$ states at $(\pi,\pi)$ do not experience any coupling to the low-energy spin degrees of freedom. A similar polaronic  phenomenology has also been employed to explain the photoemission spectra from systems ranging from ionic insulators \cite{Citrin74}, charge density wave materials (K$_{0.3}$MoO$_{3}$, (TaSe$_{4}$)$_{2}$I) \cite{Perfetti01,Perfetti02}, the Mott-Hubbard system VO$_{2}$ \cite{Okazaki04}, and the manganites \cite{Dessau98,Perebeinos00,Mannella05}. The importance of lattice trapping in the Mott insulator becomes clear when considering that photoemission introduces a positively charged hole into an insulating lattice, and in the case of Ca$_{2}$CuO$_{2}$Cl$_{2}$, one with a highly ionic CaCl rocksalt layer. Although lattice polarons have been discussed in the cuprates for some time (for instance, see Refs. \onlinecite{Alexandrov94} and \onlinecite{Zhao97}), we believe this work represents direct evidence from ARPES of lattice polaron formation in the undoped cuprates. 

In simple polaron models, the separation, $\Delta E$, between the lowest lying state (i.e. the actual QP) and the centroid of the main peak provides an estimate of the average number of phonons, $\langle n \rangle$, dressing the polaron \cite{Mahan00}. For an insulator, we take the longitudinal optical (LO) phonon branch here to be the characteristic frequency, $\omega_{0} \sim 40$ meV, yielding $\langle n \rangle = 0.42 / 0.04 \sim 10$ for the LHB. For $\langle n \rangle = 10$, the Poisson distribution is effectively indistinguishable from a Gaussian, consistent with our fits. Here we use the value of 40 meV, as this is in the range of energies for LO phonons established by both optical reflectivity and inelastic neutron scattering  in La$_{2}$CuO$_{4}$ \cite{Falck92,Pintschovius91}. In principle, choosing $\omega_{0}$ can significantly change the absolute value of the coupling constants and $\langle n \rangle$, but should not change our qualitative arguments, at least while remaining within the strong coupling regime. This surprisingly large value of $\langle n \rangle$ puts us well within the small polaron limit where the QP residue $Z \rightarrow 0$ and the photohole is completely localized. The ARPES spectra then consists of a superposition of narrow bands separated by $\omega_{0}$, which in the solid state would be blurred into a single Gaussian envelope, as we have observed. Moreover, recent detailed numerical calculations of the single-particle spectral function by R\"{o}sch \emph{et al.} which utilize a $t-J$ model framework to describe the electronic part of the system and a realistic shell model for the phonons, also suggest a small polaron-like lineshape and broadening which agree with what is observed experimentally \cite{Rosch05}.

\begin{figure}[t!]
\includegraphics{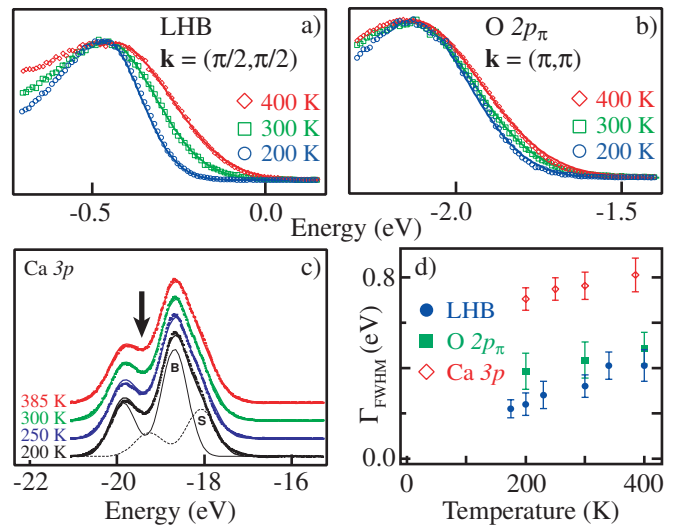} \vspace{0in}
\caption{Temperature dependence of the LHB at $(\pi / 2,\pi / 2)$ (a) and O $2p_{\pi}$ state at $(\pi,\pi)$ (b). (c) Temperature dependence of the Ca $3p$ core level, with fits to a pair of Gaussian doublets. B and S denote the individual contributions of the bulk and surface, respectively. In (a)-(c), raw data points are shown as symbols, while Gaussian fits are shown as solid lines. (d) Temperature dependence of the FWHM of the LHB (circles), O $2p_{\pi}$ (squares), and Ca $3p$ core level (diamonds).} \label{TemperatureEDCs}
\end{figure}

To further corroborate this, we measured the temperature dependence of the LHB, along with the O $2p_{\pi}$ band and the Ca $3p$ core levels, as shown in Fig. \ref{TemperatureEDCs} where the spectra have been scaled and shifted to the peak position for the purpose of comparison. The temperature dependence at the top of the LHB and the top of the O $2p_{\pi}$ band are shown with fits in Figs. \ref{TemperatureEDCs}a and \ref{TemperatureEDCs}b, respectively. $\Gamma_{\mathrm{LHB}}$ exhibits a much stronger temperature dependence than would be expected from conventional electronic interactions, which in this case would essentially unobservable, as also reported in Ref. \onlinecite{Kim02}. However, the O $2p_{\pi}$ peak also exhibits substantial temperature broadening, again suggesting that FCB is not unique to the LHB.

Data from the Ca $3p$ core level taken with He II photons are also shown in Fig. \ref{TemperatureEDCs}c after the subtraction of a Shirley-like background. Upon closer inspection, the Ca $3p$ core levels are comprised of two spin-orbit split doublets \cite{Koitzsch02}, resulting in the small shoulder at $\sim$ -17.8 eV. We ascribe the second doublet to a surface core level shift (SCLS) of Ca$^{2+}$ in the ionic CaCl cleavage plane arising from the altered Madelung potential from the surface termination. This has been observed in many materials, including other cuprates such as YBa$_{2}$Cu$_{3}$O$_{7-\delta}$ \cite {Schabel98}, and does not necessarily affect the relevant electronic states in the underlying CuO$_{2}$ plane. Through a simple analysis using the known structure and taking the escape depth as 10 \AA\ \cite{Schabel98}, we estimate the surface intensity fraction to be $S_{f} = 31 \%$, roughly consistent with experiment. Even without fitting, the raw data exhibit observable temperature broadening, as evidenced by the dip at -19.5 eV (arrow) becoming more pronounced at lower temperatures. Fits to a pair of Gaussian doublets are shown, where the parameters $S_{f}$ ($33\%$), spin-orbit splitting (1.19 eV), and the SCLS (0.61 eV) were kept fixed with temperature. The better agreement of Gaussians for the Ca $3p$ core level, as opposed to Lorentzian or Doniach-\v{S}unji\'{c} lineshapes, again suggests that lattice effects may be the dominant broadening mechanism \cite{Citrin77}.

Before proceeding further, we will rule out spurious extrinsic effects which could contribute to the observed broadening. Surface roughness was determined to be unimportant from well-defined laser reflections ($\pm 1^{\circ}$) and highly $\mathbf{k}$-dependent valence band spectra. The strong temperature and momentum dependence (later discussed) of $\Gamma$ also rule out the possibility of impurity states or electrostatic charging as significant factors. In particular, peaks become sharper at lower temperatures, opposite to charging effects which should increase as samples become more insulating. We attempted to minimize the effects of charging by freshly cleaving the bottom side of the sample before mounting in order to minimize contact resistance, as well as using the thinnest possible samples. Still, charging was typically observed below 170 K and manifested itself as a higher binding energy shift and a broadening of the spectra upon varying the incident photon flux. We cannot rule out that some inhomogeneous charging may have contributed to some of the sample-to-sample variations in $\Gamma$, represented by the error bars in Fig. \ref{TemperatureEDCs}d. However, we observed similar $d\Gamma / dT$ on multiple individual cleaves. A breakdown of the sudden approximation or extrinsic losses were ruled out by varying the kinetic energy of the photoelectrons between 9 and 27 eV, while $\Gamma_{\mathrm{LHB}}$ remained invariant.

A summary of the temperature dependence is shown in Fig. \ref{TemperatureEDCs}d. Between $200 <$ T $<400$ K, $d\Gamma / dT$ was estimated to be $1.0 \pm 0.3$ meV / K for the LHB, $0.6 \pm 0.4$ meV / K for the O $2p_{\pi}$, and $0.7 \pm 0.2$ meV / K for the Ca $3p$ core level. Core level measurements of alkali halides such as KCl, where polaron formation is also observed, give comparable values of $d\Gamma / dT$ ($\sim$ 0.8 meV / K) \cite{Citrin74}. The temperature dependence of the FCB can be explained as arising from the additional thermal population of phonons in a Bose-Einstein distribution. The absolute values of $\Gamma$ for the three states are also found to be rather different. The fact that one observes very different values of $\Gamma$ and temperature dependences for the LHB, O $2p_{\pi}$, and Ca $3p$ states raises a number of important issues. The first is that while the lattice contributes to the observed broadening of all three states, their coupling strengths to various phonon branches could potentially be very different depending on the orbital character and symmetry of the electronic wavefunctions. Therefore, the LHB, O $2p_{\pi}$, and Ca $3p$ states are likely coupled to different phonon branches with different symmetries, frequencies, and coupling strengths. This is corroborated by calculations in Ref. \onlinecite{Rosch05} which find rather different spectral distributions for the coupling of phonons to singlet states as opposed to non-bonding O $2p$ orbitals.

The second issue to be considered is the presence of additional decay channels, in addition to electron-lattice, which may also be relevant. For the O $2p_{\pi}$ and Ca $3p$ states which sit deeper in binding energy ( $\sim$ -2 eV and -19 eV, respectively), there are interband electron-hole decay channels which will contribute to the lifetime broadening. Still, the pronounced temperature dependence observed in the O $2p_{\pi}$ and Ca $3p$ linewidths in Fig. \ref{TemperatureEDCs} and good agreements to a Gaussian lineshape clearly suggest  strong electron-phonon contributions. For the case of the LHB, the creation of such electron-hole pairs is suppressed by the presence of the charge-transfer gap, and will therefore not contribute to the linewidth. Still, strong electronic / magnetic correlation effects could also potentially play an important part in contributing to the observed linewidth, in concert with electron-phonon interactions.  As shown in Fig. \ref{TemperatureEDCs}, the LHB clearly exhibits the strongest temperature broadening of all three states, and therefore some enhanced broadening could arise due to electronic effects, in combination with lattice polaron formation. Indeed, theoretical calculations indicate that in such strongly interacting systems, the electron-electron and electron-phonon interactions cannot be considered independently, but can instead feedback upon and enhance one another \cite{Mishchenko04,Macridin04}.

While we have demonstrated the importance of lattice effects to the lineshape, strong electron correlations obviously remain crucial for describing many of the global features such as the dispersion of the LHB peak which still tracks the extended $t-J$ model calculations \cite{Tohyama00}, and the transfer of spectral weight to higher energies \cite{Meinders93}. While the apparently dispersive nature of the LHB might appear contradictory to a polaronic scenario, recent theoretical works have resolved this apparent paradox. Calculations which have incorporated electron-phonon interactions into the $t-J$ framework show that while the true QP is self-trapped and has vanishing intensity, much of the remaining spectral weight forms an incoherent peak which tracks the underlying dispersion of the bare $t-J$ model \cite{Mishchenko04}, much like experiment. Other recent work has shown in general for polaronic systems that the centroid of incoherent weight tracks the dispersion predicted in the absence of electron-phonon interactions \cite{Rosch04}. This also suggests that the $t-J$ model still provides a good approximation for the dispersion of the single hole in the ``frozen'' lattice. 

\begin{figure}[t!]
\includegraphics{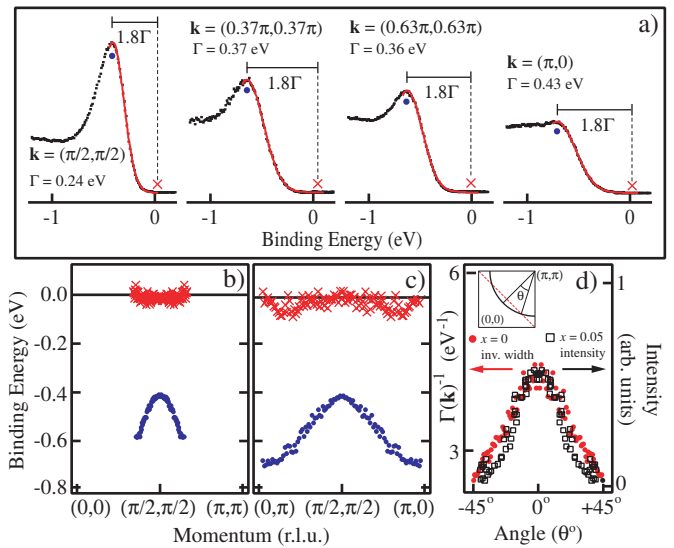} \vspace{0in}
\caption{(a) Data from the LHB taken at various $\mathbf{k}$ points, together with Gaussian fits (red). $\mathbf{k}$-dependent dispersion of the LHB, $\varepsilon(\mathbf{k})$ (blue circles), along the $(0,0)-(\pi,\pi)$ (b) and (0,$\pi$)-($\pi$,0) (c) directions. Red crosses represent $\varepsilon(\mathbf{k}) + 1.8 \, \Gamma_{\mathrm{LHB}} (\mathbf{k})$. Data in (a) (c) were taken at 200 K. (d) Comparison of the low-energy spectral weight from Ca$_{1.95}$Na$_{0.05}$CuO$_{2}$Cl$_{2}$\ with $\Gamma_{\mathrm{LHB}}(\mathbf{k})^{-1}$ as a function of angle, as defined in the inset.} \label{MomentumDependence}
\end{figure}

Our experiments have also revealed an intriguing momentum anisotropy of $\Gamma_{\mathrm{LHB}}$ which may have profound implications for the emergence of the first doped hole states. In particular, our analysis reveals that as a function of momentum, the width of the LHB peak,  $\Gamma_{\mathrm{LHB}}(\mathbf{k})$, increases directly proportional to the binding energy of the peak maximum, $\varepsilon(\mathbf{k})$, such that $\varepsilon(\mathbf{k}) = \alpha \, \Gamma_{\mathrm{LHB}}(\mathbf{k})$. Empirically, we have found that $\alpha = 1.8$, and at $1.8 \, \Gamma$ away from the peak, the intensity has decayed to $\sim 10^{-4}$ of the peak value. In Fig. \ref{MomentumDependence}a, we show data from Ca$_{2}$CuO$_{2}$Cl$_{2}$ together with Gaussian fits at various locations in $\mathbf{k}$-space, along with crosses to denote the position of $\varepsilon(\mathbf{k}) + 1.8 \, \Gamma_{\mathrm{LHB}} (\mathbf{k})$. The momentum dependence of these features along both the $(0,0)-(\pi,\pi)$ and $(0,\pi)-(\pi,0)$ directions is summarized in Figs. \ref{MomentumDependence}b and \ref{MomentumDependence}c. While the peak position changes considerably, the onset remains roughly fixed, and thus the width appears to change to accommodate the dispersion of the peak. Here we should note that the simple Gaussian fit continues to work well over an extended range in $\mathbf{k}$-space. This is somewhat surprising, given that the original assumptions for a Poisson distribution (and hence a Gaussian profile) for the spectral function were based on the idealized case of a single electronic state in a completely nondispersive band. Given that the dispersion of the LHB in the frozen lattice is $\sim$ 300 meV, this is considerably larger than the characteristic phonon frequencies, and therefore it is not clear that a Gaussian lineshape would remain a realistic approximation. Nevertheless, from an empirical standpoint, a Gaussian profile appears to describe the low-energy portion of the spectral lineshape of not only the LHB, but also the O $2p_{\pi}$ band quite well.

In a simple picture where $\Gamma$ is related to the electron-phonon coupling strength, the photohole would effectively be coupled more strongly to the lattice at $\mathbf{k}$ = $(\pi,0)$, the $d$-wave antinode, than at $\mathbf{k}$ = $(\pi / 2,\pi / 2)$, the $d$-wave node. Recent results from Ca$_{2-x}$Na$_{x}$CuO$_{2}$Cl$_{2}$ have demonstrated that spectral weight first emerges near $(\pi / 2,\pi / 2)$ upon hole doping, while only faint intensity is visible near $(\pi,0)$ \cite{Ronning03b,Shen04b}, and similar results were reported for other lightly hole doped cuprates such as La$_{2-x}$Sr$_{x}$CuO$_{4}$ and Bi$_2$Sr$_2$CaCu$_2$O$_{8+\delta}$ \cite{Yoshida03,Marshall96}. The momentum dependence of the inverse peak width in undoped Ca$_{2}$CuO$_{2}$Cl$_{2}$, $\Gamma_{\mathrm{LHB}}(\mathbf{k})^{-1}$, exhibits a qualitative similarity to the low energy spectral weight in the lightly doped materials. This is shown in Fig. \ref{MomentumDependence}d, where the spectral weight along the ostensible Fermi surface for Ca$_{1.95}$Na$_{0.05}$CuO$_{2}$Cl$_{2}$ is integrated in a narrow window around E$_{\mathrm{F}}$ (E$_{\mathrm{F}}$ $\pm 10$ meV) \cite{Shen04b} and compared with $\Gamma_{\mathrm{LHB}}(\mathbf{k})^{-1}$ along the antiferromagnetic zone boundary (dashed red) from the undoped system. The correspondence between these two quantities may suggest that the lack of well-defined QPs near $(\pi,0)$ in the lightly doped compounds may be related to the apparently anisotropic coupling in the parent compound.

In conclusion, we have established through a comparison of the LHB with the O $2p {\pi}$ and Ca $3p$ states that lattice interactions constitute a significant contribution to the observed FCB and the photohole exhibits behavior similar to a small polaron. We also find a compelling correspondence between the anisotropy of the photohole-lattice coupling and the momentum dependence of the low-lying spectral weight in the lightly hole doped cuprates. While strong correlation effects certainly dictate many important aspects, such as the existence of the Mott gap and the dispersion of spectral weight, our work demonstrates that the inclusion of lattice interactions should be important to fully explain the dynamics of the single hole in the antiferromagnetic insulator, and we view these results as complementary, and not contradictory to results from purely electronic (i.e. $t-J$ or Hubbard) models. This work emphasizes that realistic calculations of many strongly correlated systems, including poorly conducting transition metal oxides and Mott insulators, should include the effects of lattice interactions together with electronic correlations.

We would like to thank A. Damascelli, T.P. Devereaux, O. Gunnarsson, A. Mishchenko, N. Nagaosa, O. R\"{o}sch, and G.A. Sawatzky for enlightening discussions. SSRL is operated by the DOE Office of Basic Energy Science under contract DE-AC03-765F00515. K.M.S. acknowledges SGF, NSERC, and the Killam Trusts for their support. ARPES measurements were also supported by NSF DMR-0304981 and ONR N00014-98-1-0195.

\end{document}